\documentclass{article}
\usepackage{amsmath,amsbsy}
\usepackage[dvips]{graphicx}
\begin{document}
\newcommand{\eps}{\ensuremath{\varepsilon}}
\newcommand{\mi}{\ensuremath{\mathrm{i}}}
\newcommand{\me}{\mathrm{e}}
\newcommand{\nn}{\nonumber \\}
\newcommand{\bra}[1]{\langle #1 |}
\newcommand{\ket}[1]{| #1 \rangle}
\newcommand{\bigbra}[1] {\big\langle #1\big|}
\newcommand{\bigket}[1] {\big|#1\big\rangle}
\newcommand{\Bigbra}[1] {\Big\langle #1\Big|}
\newcommand{\Bigket}[1] {\Big|#1\Big\rangle}
\newcommand{\dif}{\ensuremath{\mathrm{d}}}
\newcommand{\bx}{\boldsymbol{x}}
\newcommand{\br}{\boldsymbol{r}}
\newcommand{\dagg}{^{\dag}}
\newcommand{\balpha}{\ensuremath{\boldsymbol{\alpha}}}
\newcommand{\dd}[1]{\frac{\partial}{\partial #1}}
\newcommand{\Funcder}[1]{\frac{\delta }{\delta #1}}
\newcommand{\funcder}[2]{\frac{\delta #1}{\delta #2}}
\newcommand{\Partder}[1]{\frac{\partial }{\partial #1}}
\newcommand{\partder}[2]{\frac{\partial #1}{\partial #2}}
\newcommand{\Der}[1]{\frac{\dif}{\dif#1}}
\newcommand{\der}[2]{\frac{\dif#1}{\dif#2}}
\newcommand{\ave}[1]{\langle #1\rangle}
\newcommand{\Ave}[1]{\Big\langle #1\Big\rangle}
\newcommand{\intd}[1]{\int\frac{\dif #1}{2\pi}}
\newcommand{\tr}{\mathrm{tr}}
\newcommand{\half}{{\frac{1}{2}}}
\newcommand{\halfS}{{\textstyle\frac{1}{2}\,}}
\newcommand{\dint}{\int\!\!\!\int}
\newcommand{\intbr}{\int\dif\br\,}
\newcommand{\DelT}{{\Delta\mathrm{T}}}
\newcommand{\DelW}{{\Delta\mathrm{W}}}
\newcommand{\s}{\mathrm{s}}
\renewcommand{\H}{\hat{H}}
\newcommand{\T}{\hat{T}}
\newcommand{\W}{\hat{W}}
\newcommand{\U}{\hat{U}}
\renewcommand{\v}{\hat{v}}
\newcommand{\V}{\hat{V}}
\renewcommand{\t}{\hat{t}}
\newcommand{\dintbr}{\int\!\!\!\int\dif\br_1\dif\br_2\,}
\newcommand{\vsp}{\vspace{0.5cm}}
\newcommand{\Vsp}{\vspace{1cm}}
\newcommand{\Wsp}{\vspace{2cm}}
\newcommand{\hsp}{\hspace{0.5cm}}
\newcommand{\Hsp}{\hspace{1cm}}
\newcommand{\HHsp}{\hspace{2cm}}
\newcommand{\mhsp}{\hspace{-0.5cm}}
\newcommand{\mvsp}{\vspace{-0.5cm}}
\newcommand{\mvvsp}{\vspace{-0.25cm}}
\renewcommand{\tilde}{\widetilde}
\newcommand{\nlim}{n\rightarrow\infty}
\newcommand{\q}{$\quad$}
\newcommand{\qq}{$\qquad$}
\newcommand{\ö}{\"o}
\newcommand{\ä}{\"a}
\newcommand{\rarr}{\rightarrow}
\newcommand{\lrarr}{\leftrightarrow}
\newcommand{\Rarr}{\Rightarrow}
\newcommand{\Lrarr}{\Longrightarrow}
\newcommand{\sumi}{\sum_{i=1}^N}
\newcommand{\sumj}{\sum_{j=1}^N}
\newcommand{\sumk}{\sum_{k=1}^n}
\newcommand{\suml}{\sum_{l=1}^n}
\newcommand{\textbs}[1]{\boldsymbol{#1}}
\newcommand{\bs}[1]{\boldsymbol{#1}}
\newcommand{\abs}[1]{|{#1}|}
\renewcommand{\it}{\textit}
\renewcommand{\bf}{\textbf}
\newcommand{\bfit}[1]{{\it{#1}}}
\newcommand{\bR}{\cal{R}}
\newcommand{\norm}[1]{||{#1}||}
\newcommand{\Norm}[1]{\big|\big|{#1}\big|\big|}
\newcommand{\NORM}[1]{\Big|\Big|{#1}\Big|\Big|}
\renewcommand{\intbr}{\int\dif\br}
\renewcommand{\sp}[2]{\bra{#1}{#2}\rangle}
\newcommand{\SP}[2]{\Bigbra{#1}{#2}\Big\rangle}
\newcommand{\ul}{\underline}
\newcommand{\eq}{\eqref}
\newcommand{\Fr}{Fr\'echet }
\newcommand{\Ga}{G\^ateaux }
\newcommand{\epsn}{\ensuremath{\epsilon}}
\renewcommand{\rm}{\mathrm}
\renewcommand{\cal}{\mathcal}
\newcommand{\bb}[2]{\big\{{#1}\,\big|\:{#2}\big\}}
\newcommand{\BB}[2]{\Big\{{#1}\,\Big|\:{#2}\Big\}}
\newcommand{\LL}{L1\cap L3}
\renewcommand{\S}{\cal{S}}
\newcommand{\KS}{\mathrm{KS}}
\newcommand{\HF}{\mathrm{HF}}
\newcommand{\eff}{\mathrm{eff}}
\newcommand{\xc}{\mathrm{xc}}
\newcommand{\ex}{\mathrm{ex}}
\newcommand{\R}{\mathrm{R}}
\newcommand{\occ}{\mathrm{occ}}
\newcommand{\corr}{\mathrm{corr}}
\newcommand{\ext}{\mathrm{ext}}
\newcommand{\Coul}{\mathrm{Coul}}
\newcommand{\HK}{\mathrm{HK}}
\newcommand{\TF}{\mathrm{TF}}
\newcommand{\la}{\lambda}
\newcommand{\FLL}{F_\rm{LL}}
\newcommand{\FL}{F_\rm{L}}
\newcommand{\Dif}{\ensuremath{\mathrm{D}}}
\newcommand{\dr}{\delta\rho}


\title{The locality hypothesis in density-functional theory:\\An exact theorem\vsp\\
}
\author{Ingvar Lindgren\footnote{ingvar.lindgren@fy.chalmers.se;
http://fy.chalmers.se/$\sim$f3ail} and Sten
Salomonson\footnote{f3asos@fy.chalmers.se}
\\Department of Physics, Chalmers University of Technology and the
G\"oteborg University,\\
G\"oteborg, Sweden } \maketitle
\begin{center}\normalsize Submitted to arXiv:Physics\end{center}

\begin{abstract}
\small \vsp The locality hypothesis in density-functional theory
(DFT) states that the functional derivative of the Hohenberg-Kohn
universal functional can be expressed as a local multiplicative
potential function, and this is the basis of DFT and of the
successful Kohn-Sham model. Nesbet has in several papers [Phys.
Rev. A \bf{58}, R12 (1998); \it{ibid.} A \bf{65}, 010502 (2001);
Adv. Quant. Chem, \bf{43}, 1 (2003)] claimed that this hypothesis
is in conflict with fundamental quantum physics, and as a
consequence that the Hohenberg-Kohn theory cannot be generally
valid. We have in a Comment to the Physical Review [Phys. Rev. A
\bf{67}, 056501 (2003)] commented upon these works and recently
extended the arguments [Adv. Quant. Chem. \bf{43}, 95 (2003)]. We
have shown that there is no such conflict and that the locality
hypothesis is inherently exact. In the present work we have
furthermore verified this numerically by constructing a local
Kohn-Sham potential for the $1s2s\,^3S$ state of helium that
generates the many-body electron density and shown that the
corresponding $2s$ Kohn-Sham orbital eigenvalue agrees with the
ionization energy to nine digits. Similar result is obtained with
the Hartree-Fock density. In addition to verifying the locality
hypothesis, this confirms the theorem regarding the Kohn-Sham
eigenvalue of the highest occupied orbital.
\end{abstract}
pacs: {02.30Sa, 31.15Ew, 31.15Pf}\\

\normalsize
\section{Introduction}

The \it{locality hypothesis} is the corner stone of
density-functional theory (DFT) and the basis for the widely used
Hohenberg-Kohn-Sham model~\cite{HK64,KS65}. The central issue is
whether the functional derivatives of the density functionals used
in DFT are strictly \it{local} potential functions, and whether a
Kohn-Sham model with local potential could lead to exact density
for ground states of arbitrary electronic systems.

Nesbet has in a series of papers claimed that the locality
hypothesis is in conflict with Fermi-Dirac statistics and the
exclusion [3-20]
principle 
 with the consequence that traditional DFT generally does not work
for systems with more than two electrons. This result has been
criticized by G\'{a}l~\cite{Ga00}, by Holas and
March~\cite{HM01,HM02}, as well as by us~\cite{LS03a}. We have
also recently discussed the matter further in a more comprehensive
review ~\cite{LS03b}. Nesbet has in unpublished works responded to
the our comment~\cite{Ne03d} as well as to that of Holas and
March~\cite{Ne03e}.

The first atomic DFT model was the Thomas-Fermi model, where the
electron cloud is approximated by a homogeneous electron gas. The
energy is then a functional of the electron density, $\rho(\br)$,
which for noninteracting electrons becomes~\cite{Th27,Fe28,Lieb81}
\begin{equation}
  \label{ETF}
  E_\TF[\rho]=T_\TF[\rho]+\intbr\,v(\br)\,\rho(\br),
\end{equation}
where
\begin{equation}
  \label{TF}
  T_\TF[\rho]=C\intbr\,[\rho(\br)]^{5/3}
\end{equation}
is the kinetic-energy functional and $v(\br)$ the local, external
potential. Minimizing this functional with respect to the electron
density, leads to the Euler-Lagrange equation
\begin{equation}
  \label{ELTF}
  \funcder{T_\TF}{\rho(\br)}+v(\br)=\mu,
\end{equation}
where $\mu$ is the Lagrange multiplier associated with the
normalization condition, $\intbr\,\rho(\br)=N$.

The original TF model is rather crude, and does, for instance, not
give rise to any electronic shell structure. The question raised
by Nesbet in several articles~\cite{Ne98,Ne01,Ne03} is whether
there exists \it{'an exact Thomas-Fermi model'} for noninteracting
electrons, which we interpret to imply that a functional
$T_\TF[\rho]$ exists such that the energy functional \eq{ETF}
would be exact for such systems. Nesbet claims that such a model
would be in conflict with the exclusion principle -- the TF
equation \eq{ELTF} with a single Lagrange multiplier could never
lead to electronic shell structure, which requires one parameter
for each shell.  In the Appendix we have derived the standard
Hartree-Fock equations, using a single Lagrange parameter, which
invalidates the claim that additional parameters are needed to
generate shell structure.

Since the existence of an 'exact TF model' is a direct consequence
of the validity of the HK theorem, Nesbet concludes that the
fundamental HK theory is incomplete. For the same reason the basic
work of Englisch and Englisch~\cite{EE84,EE84a} as well as most of
the current literature~\cite{PY89,DG90,Le82,Li83,GL94,PL97,Lee03}
is refuted.

There exists today an overwhelming amount of evidence for the
validity of the locality hypothesis in DFT, and it might seem
superfluous to produce another piece. Since the matter is still
under debate in the literature, however, we shall here try to
present simple arguments how the conflict has arisen and how it
can be resolved. We shall also present in our opinion strong
numerical evidence in favor of the hypothesis.

\section{Functional derivatives}
Generally, a density functional ($F[\rho]$) is defined as a
mapping of a normed space of densities\footnote{The term 'normed
space' implies that all densities of the space have a definite,
finite norm, such as $\norm{\rho(\br)}=\intbr\,\abs{\rho(\br)}$,
but the densities need not be normalized.} ($M$) on the space of
real numbers ($\bR$)~\cite{BB92},
\begin{equation}
  \label{Func}
  F[\rho]:M\rarr\bR,
\end{equation}
a mapping that has to be \it{unique} in the sense that a certain
density corresponds to a single real number.

\subsection{The \Ga derivative}
The \it{\Ga or weak differential} of a density functional
$F[\rho]$ is
defined~\cite[p.46]{BB92},~\cite[p.293]{LS61},~\cite[Eq.60]{Lee03}
\begin{equation}
  \label{Ga}
  \dif F[\rho_0,\dr]=\lim_{\la\rarr0}\frac{F[\rho_0+\la\dr]-F[\rho_0]}{\la},
\end{equation}
or alternatively
\begin{equation}
  \label{Ga1}
  \delta F[\rho_0,\la\dr]=F[\rho_0+\la\dr]-F[\rho_0]=\la\dif F[\rho_0,\dr]
  +\omega(\rho_0,\la\dr),
\end{equation}
where
\begin{equation}
  \label{omegaG}
  \lim_{\la\rarr0}\frac{\omega(\rho_0,\la\dr)}{\la}=0.
\end{equation}
In principle, this differential depends on the direction $\dr$,
and it need neither be linear nor continuous in $\dr$. Generally,
it can be expressed
\begin{equation}
  \label{Diff1}
  \dif F[\rho_0,\delta\rho]=
  \intbr\,F'\big([\rho_0,\dr],\br\big)\,\delta\rho(\br),
\end{equation}
where $F'\big([\rho_0,\dr],\br)$ is a function of $\br$ -- not
necessarily continuous -- and a functional of $\rho$ as well as of
$\dr$. If the differential is \it{linear and continuous} in $\dr$,
then the function must be independent of $\dr$, and the
differential can be expressed
\begin{equation}
  \label{Diff}
  \dif F[\rho_0,\delta\rho]=
  \intbr\,F'\big([\rho_0],\br\big)\,\delta\rho(\br).
\end{equation}
$F'\big([\rho_0],\br\big)$ is here a local, finite, multiplicative
function of $\br$ that is single-valued for given $\rho_0$, and
conventionally referred to as the \it{\Ga derivative} of the
functional $F$ at the density
$\rho_0$~\cite[p.46]{BB92},~\cite[Eq.61]{Lee03},
\begin{equation}
  \label{Ga3}
  \Big(\funcder{F[\rho]}{\rho(\br)}\Big)_{\rho=\rho_0}=F'\big([\rho_0],\br\big).
\end{equation}

\subsection{The \Fr derivative}
The \it{\Fr or strong differential} $\dif F[\rho_0,\dr]$ can be
defined~\cite[p.37]{BB92},~\cite[p.292]{LS61}
\begin{equation}
  \label{Fr}
  \delta F[\rho_0,\dr]=F[\rho_0+\dr]-F[\rho_0]=
  \dif F[\rho_0,\dr]+\omega(\rho_0,\dr),
\end{equation}
where
\begin{equation}
  \label{omega}
  \lim_{\norm{\dr}\rarr0}\frac{\omega(\rho_0,\dr)}{\norm{\dr}}=0.
\end{equation}
Here, the limit has to be \it{uniform} in the neighborhood of the
density $\rho_0$. The \Fr differential can also be expressed in
the form \eq{Diff} with the derivative in the form of a local
potential function~\cite[p.41]{BB92}.

The criterium for \Fr differentiability \eq{omega} is stronger
than the corresponding \Ga criterium \eq{omegaG}. If the \Ga
derivative exists in the neighborhood of a density and is
\it{uniformly continuous} in this neighborhood, then the \Ga
derivative at that density is also a \it{\Fr
derivative}~\cite[p.47]{BB92},~\cite[p.295]{LS61}. This requires
that the functional is defined also for unnormalized densities. In
DFT it is possible to stay within the normalization domain, and
then \Ga differentiability is sufficient.

In the next section we shall summarize the basics of the Kohn-Sham
model as a background of the locality analysis. (For further
details regarding the basic theory, see, e.g.,
refs.~\cite{PY89,DG90,JG89,Lee03}.)

\section{The Hohenberg-Kohn-Sham model}
\subsection{The Hohenberg-Kohn theorem}
The Hohenberg-Kohn (HK) universal functional~\cite{HK64} is in the
constrained-search formulation given
by~\cite{Le79,Li83}~\footnote{This is known as the \it{Levy-Lieb
functional} and sometimes denoted $\FLL[\rho]$.}
\begin{equation}
  \label{FHK}
  F_\HK[\rho]=\min_{\Psi\rightarrow
    \rho}\bigbra{\Psi}\T+\W\bigket{\Psi}=T_\HK[\rho]+W_\HK[\rho].
\end{equation}
Here, $\T$ and $\W$ represent the kinetic-energy and
electron-electron-interaction operators of the $N$-electron
system, respectively,
\begin{equation}
  \label{TW}
  \T=-\sum_{i=1}^N\halfS\nabla2_i\,; \quad   \W=\sum_{i<j}^N
\frac{1}{|\br_i-\br_j|}.
\end{equation}
The normalized wavefunction $\Psi$ belongs to the \it{Sobolev
space} $H1(\bR^{3N})$~\cite{Le79,Li83,LS03b,Lee03}, and the
corresponding functional is defined for all \it{N-representable}
densities~\cite{DG90}.

The HK energy functional of a system with an external, local
potential $v(\br)$ is
\begin{equation}
  \label{EHK}
  E_v[\rho]=F_\HK[\rho]+\int\dif\br\,\rho(\br)\,v(\br),
\end{equation}
and the energy, $E_{v0}$, and the electron density, $\rho_0(\br)$,
of the ground-state (or the lowest eigenstate of certain
symmetry~\cite{GL76}) are obtained by minimizing this functional
over normalized densities~\cite{DG90}
\begin{equation}
  \label{MinEHK}
  E_{v0}=\min_{\rho\rarr N}E_v[\rho]=E_v[\rho_0].
\end{equation}

As in our previous works~\cite{LS03a,LS03b}, we shall extend the
definition of the HK functional \eq{FHK} in a straightforward way
by allowing the wavefunctions to vary also outside the normalized
domain of the Sobolev space, which is needed in order to be able
to apply the Euler-Lagrange procedure. The minimization then leads
to the Euler-Lagrange equation
\begin{equation}
  \label{ELHK}
  \funcder{F_\HK[\rho]}{\rho(\br)}+v(\br)=\mu,
\end{equation}
where $\mu$ is the Lagrange parameter for the normalization
constraint, $\intbr\,\rho(\br)=N$.

If we omit the electron-interaction part, $\W$, of the HK
functional \eq{FHK}, then the EL equation \eq{ELHK} is quite
analogous to the Thomas-Fermi equation for noninteracting
electrons \eq{ELTF}. The kinetic part, $T_\HK[\rho]$, of the HK
functional \eq{FHK} is then the kinetic-energy functional of the
\it{'exact Thomas-Fermi model'}, which thus is a direct
consequence of the HK theorem.

\subsection{The Kohn-Sham model}
In the Kohn-Sham (KS) model, the interacting system is replaced by
a system of \it{noninteracting electrons}, moving in the local
Kohn-Sham potential, $v_\KS(\br)$,
\begin{equation}
  \label{KSeq}
\big[
-\halfS\nabla2+v_\KS(\br)\big]\phi_i(\br)=\eps_{i}\,\phi_i(\br).
\end{equation}
The analogue of the energy functional \eq{EHK} is then
\begin{equation}
  \label{EKS}
  E_v[\rho]=T_\s[\rho]+\int\dif\br\,\rho(\br)\,v_\KS(\br),
\end{equation}
where
\begin{equation}
  \label{TKS}
   T_\s[\rho]=\min_{\Phi\rightarrow\rho}\bigbra{\Phi}\T
  \bigket{\Phi},
\end{equation}
and $\Phi$ is a single Slater-determinantal wavefunction in the
same functional space as previously. Minimizing the energy
functional, leads in analogy with the Euler-Lagrange equation
\eq{ELHK} to
\begin{equation}
  \label{MinKE}
  \funcder{T_\s[\rho]}{\rho(\br)}+v_\KS(\br)=\mu.
\end{equation}
The two equations lead to the same solution, if -- apart from a
constant --
\begin{equation}
  \label{VKS}
  v_\KS(\br)=\funcder{F_\HK[\rho]}{\rho(\br)}-\funcder{T_\s[\rho]}{\rho(\br)}+v(\br).
\end{equation}
If this potential were known, solving the KS equations
self-consistently would, in principle, yield the exact electron
density and, if the HK functional \eq{EHK} were known, the exact
total energy of the ground state -- or the lowest state of a given
symmetry. In addition, it has been shown that the eigenvalue of
the highest occupied KS orbital yields the exact ionization
energy~\cite{PPL82,AP84,AB85,PL97}.

A crucial point in the KS procedure is that the functionals
involved are differentiable and that the functional derivative is
in the form of a \it{local} potential function. This is now a
well-established fact. In the standard DFT procedure, the
variations are restricted to normalized densities, and then \Ga
differentiability is sufficient. The \it{\Ga differentiability} of
the functional \eq{FHK} was rigorously demonstrated two decades
ago by Englisch and Englisch~\cite{EE84,EE84a}, based upon works
of Lieb~\cite{Li83}, and the derivative was shown to be in the
form of a \it{local} potential function.\footnote{The proof
concerns the so-called Lieb functional, usually denoted
$\FL[\rho]$, which in the case of nondegenerate ground states we
are concerned with here is identical to the functional \eq{FHK}.}
These results have recently been carefully confirmed by van
Leeuwen~\cite{Lee03}. By extending the definitions to unnormalized
densities, as mentioned above, we have demonstrated that the
functionals involved have a \Ga derivative according to the
definition \eq{Diff} above also in the extended domain. This
derivative is -- at least for practical purposes -- also a \it{\Fr
derivative}~\cite{LS03a,LS03b,LS04}.\footnote{This holds for
calculations with finite basis set, where the kinetic energy has a
finite upper bound, but it may not be rigorously true when the
number of dimensions is infinite. (See ref.~\cite{LS03b} for more
detailed discussion).}

\section{Nesbet's locality dilemma and its solution}
\label{Dil}
\subsection{The problem}
In his efforts to demonstrate that standard DFT is in conflict
with fundamental physics, Nesbet considers the ground state -- or
lowest state of given symmetry -- of a system of noninteracting
electrons. For a two-electron system this can be represented by a
determinantal wavefunction $\Phi(\br_1,\br_2)$, composed of
spin-orbitals with the space part satisfying the single-electron
Schr\ödinger equation
\begin{equation}
  \label{Schr}
  \big(\t+v(\br)\big)\phi_i(\br)=\eps_i\,\phi_i(\br)
\end{equation}
with $\t$ being the kinetic-energy operator $\t=-\half\nabla^2$
and $v(\br)$ a local external potential. For the kinetic energy
and electron density of the system Nesbet uses the expressions
\begin{eqnarray}
  \label{TN}
  T&=&\bra{\phi_1}\t\ket{\phi_1}+\bra{\phi_2}\t\ket{\phi_2}\\
  \label{DensN}
  \rho(\br)&=&\rho_1(\br)+\rho_2(\br)=|\phi_1(\br)|^2+ |\phi_2(\br)|^2,
  \end{eqnarray}
  valid for normalized orbitals.
By making small orbital changes $\delta\phi_i$, the kinetic energy
of the ground state is to leading order changed by (Eq. 9
in~\cite{Ne01})
\begin{equation}
  \label{DeltaTN}
  \dif
  T=\intbr\,\big(\eps_1-v(\br)\big)\,\delta\rho_1(\br)+
  \intbr\,\big(\eps_2-v(\br)\big)\,\delta\rho_2(\br),
\end{equation}
where
\[\delta\rho_i(\br)=\delta\phi_i^*(\br)\phi_i(\br)+c.c.\]
This is neither of the form \eq{Diff} nor of the form \eq{Diff1}
outside the domain of normalized orbitals (where
$\intbr\,\delta\rho_i(\br)\neq0$) when the eigenvalues are
different. Therefore, $\dif T$ cannot be a differential of a
density functional in this region, and no density-functional
derivative exists, as we have demonstrated before~\cite{LS03a}.
Instead, Nesbet introduces \it{'orbital-dependent'} derivatives
\begin{equation}
  \label{DerivN}
  \funcder{T}{\rho_i(\br)}=\eps_i-v(\br)
\end{equation}
and interprets them as \it{\Ga derivatives}, which is obviously
neither in accordance with the standard definition \eq{Diff}
above, nor with the extended definition \eq{Diff1}.

In spite of the fact that $T$ is not a density functional in the
extended domain, Nesbet applies the chain rule,
\begin{equation}
  \label{Chain}
  \funcder{T}{\rho_i(\br)}=\funcder{T}{\rho(\br)}\:
   \funcder{\rho(\br)}{\rho_i(\br)}=\funcder{T}{\rho(\br)},
\end{equation}
which together with the result \eq{DerivN} leads to a conflict
with the TF equation \eq{ELTF}, if the orbital energies are
different,
\begin{equation}
  \label{Conflict}
  \eps_1\neq\eps_2 \; \Rarr
  \;\funcder{T}{\rho_1(\br)}\neq\funcder{T}{\rho_2(\br)}.
\end{equation}
The reason for the conflict is that the kinetic-energy expression
\eq{TN}, which obviously is an \it{orbital functional}, is \it{not
a density functional} outside the domain of normalized orbitals
when the eigenvalues are different. Therefore, the chain rule
\eq{Chain} cannot be used in that region, and the conflict is only
apparent, as we have demonstrated before~\cite{LS03a,LS03b} and
emphasized also by Holas and March~\cite{HM01}. We shall now
indicate how this conflict can be resolved.

\subsection{The solution}
In the treatment of Nesbet the density variations are allowed to
go outside the normalization domain. At the same time it is
emphasized that the kinetic-energy expression should be an orbital
as well as a density functional, which, of course, has to be the
case in the extended domain.

The kinetic energy and the electron density are not \it{a priori}
defined outside the normalization domain, and we have to choose a
proper extension. A natural extension is simply to use the
standard expressions also outside the
normalization~\cite{LS03a,LS03b},
\begin{eqnarray}
  \label{TDet}
  &&\mhsp T=\dintbr\,\Phi^*(\br_1,\br_2)(\t_1+\t_2)\Phi(\br_1,\br_2)=
  \bra{\phi_1}\t\ket{\phi_1}\,\bra{\phi_2}\phi_2\rangle+\bra{\phi_2}\t\ket{\phi_2}\,
  \bra{\phi_1}\phi_1\rangle\\\label{RhoDet}
  &&\mhsp \rho(\br)=2\intbr_2\,\abs{\Phi(\br,\br_2)}^2=|\phi_1(\br)|^2\,\bra{\phi_2}\phi_2\rangle+
 |\phi_2(\br)|^2\,\bra{\phi_1}\phi_1\rangle,
\end{eqnarray}
assuming the orbitals to be mutually orthogonal. Since the
orbitals are not normalized, it is vital to maintain the
normalization integrals in the expression to be varied. Instead of
the expression \eq{DeltaTN}, the differential of the kinetic
energy due to orbital changes now becomes~\cite{LS03a}
\begin{equation}
  \label{DeltaT}
  \dif
  T=\int\dif\br\,\Big(\frac{\eps_1+\eps_2}{2}-v(\br)\Big)\,\delta\rho(\br),
\end{equation}
which is in accordance with the expression \eq{Diff} also for
variations outside the normalization. This verifies that the form
\eq{TDet} is a unique functional of the total density \eq{RhoDet}
in the neighborhood of the ground-state density, and it yields the
derivative
\begin{equation}
\label{Deriv}
  \funcder{T}{\rho(\br)}=\frac{\eps_1+\eps_2}{2}-v(\br).
\end{equation}
This is in contrast to the result of Nesbet \eq{Conflict}
\it{orbital independent}. The derivative is the same in all
directions -- also outside the normalization -- and is, as
discussed above, of \it{\Fr type}.

We can now identify the constant in the derivative \eq{Deriv} with
the Lagrange multiplier in the 'Thomas-Fermi' equation \eq{ELTF},
\begin{equation}
\label{Lagr}
  \mu=\frac{\eps_1+\eps_2}{2}.
\end{equation}
Thus, there is only a single Lagrange parameter, when the
extension to the unnormalized domain is done in a consistent way,
and \bfit{the conflict \eq{Conflict} with the Thomas-Fermi
equation is removed}.

The crucial point in the arguments is the chain rule \eq{Chain}.
The more exact form of this rule is
\begin{equation}
  \label{Chain2}
  \funcder{T}{\phi^*_i(\br)}=\intbr'\funcder{T}{\rho(\br')}\:
   \funcder{\rho(\br')}{\phi^*_i(\br)}.
\end{equation}
If we assume that the density is given by \eq{DensN}, then
$\funcder{\rho(\br')}{\phi^*_i(\br)}=\delta(\br-\br')\,\phi_i(\br)$
and
$\funcder{T}{\phi^*_i(\br)}=\funcder{T}{\rho(\br)}\,\phi_i(\br)$,
which is equivalent to the simplified chain rule \eq{Chain}. But
this holds only inside the normalization domain, where the density
can be written in the form \eq{DensN}. In the treatment of Nesbet,
however, the variations have to go outside this domain, otherwise
the orbital dependence of the differential \eq{DeltaTN} would
vanish, and there would be no conflict.

With the expression \eq{RhoDet} for the density, valid also
outside the normalization domain, the correct chain rule
\eq{Chain2} yields
\begin{eqnarray}
  \funcder{T}{\phi_1^*(\br)}=
  \Bigg[\funcder{T}{\rho(\br)}\,\sp{\phi_2}{\phi_2}+
  \Bigbra{\phi_2}\funcder{T}{\rho}\Bigket{\phi_2}\Bigg]\,
  {\phi_1(\br)}.
 \end{eqnarray}
This is compared with the direct orbital derivation~\cite{LS03a}
\begin{equation}
  \label{OrbDer}
  \funcder{T}{\phi_1^*(\br)}=\Big[\big(\eps_1+\eps_2-v(\br)\big)\sp{\phi_2}{\phi_2}-
  \bra{\phi_2}v\ket{\phi_2}\Big]\,\phi_1(\br),
\end{equation}
and identification yields the same expression for the derivative
as before \eq{Deriv}. \it{Thus, by using proper expressions and
using the chain rule in a proper way, the conflict observed by
Nesbet between direct orbital derivation and the chain rule has
disappeared.}

\subsection{Comments}

The basic claim of Nesbet is that \it{'density functional
derivatives of the Hohenberg-Kohn universal functional cannot be
equivalent to local potential functions, as assumed in established
literature~\cite{PY89,DG90}, for more than two electrons in a
compact electronic system'}~\cite{Ne03a}, and it is concluded that
\it{'the TF theory is not equivalent to Kohn-Sham theory, even if
the exact Hohenberg-Kohn universal functional were known and
used'}~\cite{Ne98}.

Reference is made to the 'well-known failure' of Thomas-Fermi
theory to describe the atomic shell structure. This failure is
unknown to us -- unless, of course, it refers to the trivial case
of the original approximation \eq{TF} from the 1920's. According
to the HK theorem, an exact kinetic-energy functional does exist
($T_\HK[\rho]$ in Eq.\eq{FHK}), but to our knowledge no
approximation to the functional beyond the original approximation
has been constructed and tested.

A related question addressed by Nesbet is that the Euler-Lagrange
equation, like the TF equation \eq{ELTF} or the HK equation
\eq{ELHK}, with a single Lagrange parameter cannot lead to
electronic shell structure. In a recent paper~\cite{Ne03f} it is
stated that \it{'the exclusion principle requires independent
  normalization of the orbital partial densities'}. Without such independent
normalization Nesbet claims that the electronic wavefunction will
collapse to the lowest single-electron state.

In the Appendix we have given an elementary counterargument by
deriving the standard Hartree-Fock equations for a two-electron
system with a single parameter for the normalization of the
\it{total} wave function. This obviously leads to shell structure,
and there is no collapse into the lowest electronic
state~\cite{Ne03d}. Parameters for normalization of individual
shells are not needed, if orbital overlap integrals are included
in the expression to be minimized.

The standard way of deriving the HF equations is, of course, to
introduce Lagrange multipliers for the orthonormality condition.
Here, however, we want to emphasize that there exist two
equivalent ways of performing a minimization by means of orbital
variations, \it{either} by enforcing orthonormality by means of
Lagrange multipliers, omitting orbital normalization and overlap
integrals, \it{or} by maintaining these integrals together with a
single Lagrange parameter for the overall normalization (and
performing the orthonormalization afterwards, if desired). In our
DFT work we used the latter method.

The reason for the failure of standard DFT is according to Nesbet
that the treatment is restricted to normalized
densities~\cite{Ne03a,Ne03b}. The Euler-Lagrange procedure
requires the functional to be defined also in infinitesimal
neighborhoods of such densities. This is regarded as \it{'a
crucial limitation of the Hohenberg-Kohn theory.'} It is assumed
that the normalization constraint 'hides' the additional
parameters needed to generate shell structure. The conclusion is
that a correct theory can only be developed using functional
derivatives of \it{orbital} densities with \it{nonlocal}
potentials.

We have shown~\cite{LS03a,LS03b} that restricting the treatment to
normalized densities constitutes no limitation. Extending the
treatment to the unnormalized domain leads to identical results --
no 'hidden' parameters are uncovered.

That DFT is complete without any additional parameters has also
been demonstrated among others by Perdew and Levy~\cite{PL97}, who
emphasize that \it{'the exact functionals themselves (and not
extraneous constraints) are responsible for the shell structure'}.

\it{Our analysis reconfirms the well-established result that DFT
with local potential is inherently exact}. The same conclusion,
using different lines of arguments, is reached by Holas and March
in their recent analysis~\cite{HM02}.

\section{Numerical demonstration of locality}
As a further demonstration of the validity of the locality
hypothesis, we have, using the method of Baerends and van
Leeuwen~\cite{BL94}, constructed numerically the local Kohn-Sham
potential for the $1s2s\,^3S$ state of neutral helium -- a state
often used by Nesbet to demonstrate the break-down of this
principle. Our starting point is the density generated by an
accurate many-body wavefunction, obtained by means of our
all-order pair procedure~\cite{SO89a}. The potential is
constructed so that the Kohn-Sham orbitals generate the many-body
density very accurately. The energy eigenvalue of the $2s$
Kohn-Sham orbital is then found to agree with the many-body
ionization energy within our numerical accuracy of nine
digits~\cite{SML04}. \it{This result verifies that a single local
potential can generate essentially exact results, also for
electrons with different energy eigenvalues, and that there is no
conflict between the locality hypothesis and the exclusion
principle,} as claimed by Nesbet. Related results (with less
numerical accuracy) have been obtained by many groups in the
past~\cite{AP84,HS89,WP93,ZMP94,HM95,GLB96,THG97,GU98,Ha98}. In
addition to verifying the locality hypothesis, our result
represents another numerical verification of the theorem regarding
the highest occupied Kohn-Sham
eigenvalue~\cite{PPL82,AP84,AB85,PL97,Ha99}.

We have obtained similar result with the Hartree-Fock density,
which also can be reproduced by a local Kohn-Sham potential. The
eigenvalue of the KS 2$s$ orbital then agrees with the HF value
with high accuracy. This demonstrates that it is possible to
represent exchange -- in the DFT sense -- by a local potential.

The latter result is in sharp contrast to the conclusions drawn,
for instance, by Nesbet and Colle~\cite{NC99,NC99a,Ne01b,Ne03},
who conclude that the Kohn-Sham  model with a local potential can
never be exact and, in particular, that a local potential can
never be better than the optimized-potential method of Talman and
Shadwick~\cite{TS76}. This potential in turn yields higher
ground-state energy than the Hartree-Fock model, as measured by
the expectation value of the Hamiltonian, $\ave{H}$, and this is
supposed to confirm the failure of the locality hypothesis for
exchange. A comparison of this kind is, of course, of no value,
since the energy in DFT is given by the HK energy functional --
not by $\ave{H}$. Needless  to say, though, that a local potential
can never in all respects reproduce the nonlocal exchange.

Our results provide counterexamples to Nesbet's conclusions drawn
in several papers~\cite{Ne03,Ne03b} that a correct theory can only
be developed using \it{orbital}-functional theory  with
\it{nonlocal} potentials. We have demonstrated that
density-functional theory with local potential is exact -- but, of
course, not identical to orbital-functional theory with nonlocal
potential.

\section{Summary}
\bfit{In summary, we have reconfirmed that the locality hypothesis
is an exact statement. The conflict with the exclusion principle,
claimed by Nesbet, is only apparent and is eliminated by using
expressions that are density functionals also outside the orbital
normalization. Our conclusion is supported by accurate numerical
results.}

\section*{Acknowledgments}
We are grateful to our student Fredrik M\öller for performing the
accurate Kohn-Sham calculations, demonstrating the validity of the
locality hypothesis, and allowing us to quote his results prior to
publication. We also acknowledge communication with Erkki
Br\ändas, Andreas G\örling, Andrew Holas, Robert van Leeuwen, and
Sam Trickey. The work has been supported by the Swedish Research
Council.

\section*{Appendix: Fermi-Dirac Statistics with a single Lagrange
Multiplier}

In this appendix we want to demonstrate in an elementary way that
it is possible to satisfy the conditions for Fermi-Dirac
statistics in a variational procedure with only a single Lagrange
multiplier. We shall do that by minimizing the expectation value
of the Hamiltonian $\bra{\Psi}\H\ket{\Psi}$ under the condition
that the total wavefunction $\Psi$ is normalized,
$\sp{\Psi}{\Psi}=1$, and show that this leads to the standard
Hartree-Fock equations. No additional parameters are needed.
Fermi-Dirac statistics is enforced by demanding that the
two-electron wavefunction is antisymmetric.

Generally, the solution of the problem is found by demanding that
\begin{equation}
  \label{E1}
  L[\Psi]=\bra{\Psi}\H\ket{\Psi}-E\sp{\Psi}{\Psi}
\end{equation}
is stationary with $E$ being the Lagrange multiplier for the
normalization of $\Psi$. This leads to the Schr\ödinger equation
\begin{eqnarray}
  \label{E2A}
  \H\Psi=E\Psi\,;\qquad
    \sp{\Psi}{\Psi}=1.
\end{eqnarray}

We consider now a two-electron system in a state where the
electrons are in different orbitals, such as the lowest triplet
state, $1s2s\,^{3}S$. The antisymmetric wavefunction can be
expressed in terms of spin-orbitals $\phi_a$ and $\phi_b$,
\begin{equation}
  \label{E3}
  \Psi_{\HF}(1,2)={\frac{1}{\sqrt{2}}}
\Big(\phi_a(1)\phi_b(2)-\phi_b(1)\phi_a(2)\Big).
\end{equation}
The hamiltonian $\H$ is here
\begin{subequations}
\begin{equation}
  \label{E4}
  \H=h_0(1)+h_0(2)+\frac{1}{r_{12}}
\end{equation}
with
\begin{equation}
  \label{E5}
  h_0=-\half\nabla^2-\frac{Z}{r}.
\end{equation}
\end{subequations}
Using Eq. \eq{E3}, the functional \eq{E1} now becomes an orbital
functional
\begin{eqnarray}
  \label{E6}
  L[\Psi_{\HF}]=L[\phi_a,\phi_b]&=&\bra{a}h_0\ket{a}\sp{b}{b}
                   + \sp{a}{a}\bra{b}h_0\ket{b}
                   - \bra{a}h_0\ket{b}\sp{b}{a}
                   - \sp{a}{b}\bra{b}h_0\ket{a}\nn
                   &+& \bra{ab }\frac{1}{r_{12}}\ket{ab}
                   - \bra{ab}\frac{1}{r_{12}}\ket{ba}
                   - E\Big(\sp{a}{a}\sp{b}{b}
                       - \sp{a}{b}\sp{b}{a}\Big),
\end{eqnarray}
which should be stationary under independent small variations of
the orbitals. So, the conditions $\delta L / \delta \phi^*_a=0$
and $\delta L / \delta \phi^*_b=0$ lead to the orbital equations
\begin{subequations}
\begin{eqnarray}
  \label{E7A}
    h_0\ket{a}\sp{b}{b}
 &+& \ket{a}\bra{b}h_0\ket{b}
  - h_0\ket{b}\sp{b}{a}
  - \ket{b}\bra{b}h_0\ket{a}\nn
 &+& \bra{b}\frac{1}{r_{12}}\ket{b}\ket{a}
  - \bra{b}\frac{1}{r_{12}}\ket{a}\ket{b}
  - E\Big(\ket{a}\sp{b}{b} - \ket{b}\sp{b}{a}\Big)
  = 0
\\
  \label{E7B}
    h_0\ket{b}\sp{a}{a}
 &+& \ket{b}\bra{a}h_0\ket{a}
  - h_0\ket{a}\sp{a}{b}
  - \ket{a}\bra{a}h_0\ket{b}\nn
 &+& \bra{a}\frac{1}{r_{12}}\ket{a}\ket{b}
  - \bra{a}\frac{1}{r_{12}}\ket{b}\ket{a}
  - E\Big(\ket{b}\sp{a}{a} - \ket{a}\sp{a}{b}\Big)
  = 0,
\end{eqnarray}
\end{subequations}
while $\delta L / \delta\phi_a=0$ and $\delta L / \delta\phi_b=0$
lead to equivalent complex conjugate equations.

Although the normalized two-electron wave function is unique (up
to a phase factor of modulus one), the orbitals are not. They need
not be individually normalized and/or orthogonal. If we have found
orbitals that satisfy the equations above, however, we can easily
transform them to become orthonormal without changing the
two-electron wavefunction. Denoting the transformed orbitals again
as $\ket{a}$ and $\ket{b}$ for brevity, we insert their properties
\begin{subequations}
\begin{eqnarray}
  \label{E9C}
  \bra{a}a\rangle=\bra{b}b\rangle&=&1
\\
  \label{E9D}
  \bra{a}b\rangle=\bra{b}a\rangle&=&0
\end{eqnarray}
\end{subequations}
into the equations \eq{E7A} and \eq{E7B} to obtain
\begin{subequations}
\begin{eqnarray}
  \label{E8A}
    h_0\ket{a}
 &+& \ket{a}\bra{b}h_0\ket{b}
  - \ket{b}\bra{b}h_0\ket{a}\nn
 &+& \bra{b}\frac{1}{r_{12}}\ket{b}\ket{a}
  - \bra{b}\frac{1}{r_{12}}\ket{a}\ket{b}
  - E\ket{a}
  = 0
\\
  \label{E8B}
    h_0\ket{b}
 &+& \ket{b}\bra{a}h_0\ket{a}
  - \ket{a}\bra{a}h_0\ket{b}\nn
 &+& \bra{a}\frac{1}{r_{12}}\ket{a}\ket{b}
  - \bra{a}\frac{1}{r_{12}}\ket{b}\ket{a}
  - E\ket{b}
  = 0
\end{eqnarray}
\end{subequations}
which are equivalent to the familiar HF equations
\begin{subequations}
\begin{eqnarray}
  \label{E9A}
    h_0\ket{a}
  + \bra{b}\frac{1}{r_{12}}\ket{b}\ket{a}
  - \bra{b}\frac{1}{r_{12}}\ket{a}\ket{b}
 &=& \eps_a\ket{a} + \eps_{ba}\ket{b}
\\
  \label{E9B}
    h_0\ket{b}
  + \bra{a}\frac{1}{r_{12}}\ket{a}\ket{b}
  - \bra{a}\frac{1}{r_{12}}\ket{b}\ket{a}
 &=& \eps_b\ket{b} + \eps_{ab}\ket{a}
\end{eqnarray}
\end{subequations}
if we denote some combinations of constants as
\begin{subequations}
\begin{eqnarray}
  \label{EA}
  \eps_a&=&E - \bra{b}h_0\ket{b}
\\
  \label{EB}
  \eps_b&=&E - \bra{a}h_0\ket{a}
\\
  \label{EAB}
  \eps_{ab} &=& \bra{a}h_0\ket{b} = \eps^*_{ba}.
\end{eqnarray}
\end{subequations}

By projecting \eq{E8A} onto $\bra{a}$ or \eq{E8B} onto $\bra{b}$,
the value of the Lagrange multiplier $E$ becomes
\begin{eqnarray}
  \label{E10}
  E = \bra{a}h_0\ket{a}
    + \bra{b}h_0\ket{b}
    + \bra{ab}\frac{1}{r_{12}}\ket{ab}
    - \bra{ab}\frac{1}{r_{12}}\ket{ba},
\end{eqnarray}
which shows that $E$ corresponds to the total energy,
$E=\bra{\Psi_{\HF}}\H\ket{\Psi_{\HF}}$. By inserting this value
into \eq{EA} and \eq{EB} we also get the orbital energies
$\eps_a$, $\eps_b$
\begin{subequations}
\begin{eqnarray}
  \label{E11}
  \eps_a &=&
           \bra{a}h_0\ket{a}
         + \bra{ab}\frac{1}{r_{12}}\ket{ab}
         - \bra{ab}\frac{1}{r_{12}}\ket{ba}
\\
  \label{E12}
  \eps_b &=&
           \bra{b}h_0\ket{b}
         + \bra{ab}\frac{1}{r_{12}}\ket{ab}
         - \bra{ab}\frac{1}{r_{12}}\ket{ba}
\end{eqnarray}
\end{subequations}
An alternative form of the total energy \eq{E10} is therefore
\begin{eqnarray}
  \label{E14}
  E &=& \eps_a + \eps_b
    - \Big( \bra{ab}\frac{1}{r_{12}}\ket{ab}
           - \bra{ab}\frac{1}{r_{12}}\ket{ba} \Big).
\end{eqnarray}

Note that it is not possible to introduce the simplification, i.e.
orbitals transformed into orthonormal orbitals, already in Eq.
\eq{E6}, \it{before} the variation. That would not lead to
meaningful orbital equations. It is the variation of the overlap
integrals that is responsible for making the orbital energies
$\eps_a$ and $\eps_b$ different.

\bibliographystyle{prsty}

\end{document}